\documentclass[aps,preprint,showpacs,showkeys,groupedaddress]{revtex4-1}
\pdfoutput=1
\usepackage{amsmath,amsthm,amssymb}
\usepackage{array}
\usepackage{graphicx}
\usepackage{subfigure}
\usepackage{fancyhdr}
\usepackage{color}
\usepackage{extarrows}
\usepackage{enumerate}
\usepackage{hyperref}

\begin{document}

\title{Beyond IT\^{O} \textit{vs.} STRATONOVICH}

\author{Ruoshi Yuan}
\affiliation{Department of Computer Science and Engineering Shanghai Jiao Tong University, Shanghai, 200240, China}
\author{Ping Ao}
\email{aoping@sjtu.edu.cn}
\affiliation{Shanghai Center for Systems Biomedicine and Department of Physics\\Shanghai Jiao Tong University, Shanghai, 200240, China}

\begin{abstract}
Recently, a novel framework to handle stochastic processes emerges from a series of studies in biology,
showing situations beyond ``It\^o \textit{vs.} Stratonovich".
Its internal consistency can be demonstrated via the zero mass limit of a generalized Klein-Kramers equation.
Moreover, the connection to other integrations becomes evident:
The obtained Fokker-Planck equation defines a new type of stochastic calculus that in general differs from the $\alpha$-type interpretation.
A unique advantage of this new approach is a natural correspondence between stochastic and deterministic dynamics,
which is useful or may even be essential in practice.
The core of the framework is a transformation from a usual Langevin equation to a form that contains a potential function with two additional dynamical matrices,
which reveals an underlying symplectic structure.
The framework has a direct physical meaning and a straightforward experimental realization.
A recent experiment offers a first empirical validation of such new stochastic integration. [This article has been published in \href{http://iopscience.iop.org/1742-5468/2012/07/P07010/}{\textit{J. Stat. Mech.} (2012) P07010}. Note that a typo in Sect.~III B has been corrected (See also [39]). ]
\end{abstract}

\date{\today}
\maketitle

\section{Introduction}
It\^{o}-Stratonovich dilemma \cite{Gardiner2004,Kampen2007,risken1996fokker}, on choosing the appropriate calculus
when integrating stochastic differential equations (SDEs) ``attracted considerable attention in the physics community"
and ``is still as elusive as ever" \cite{McClintock}. Recent explorations of SDEs \cite{Zhu2004,Ao2004,ao2005laws,ao2005metabolic,Kwon2005,Yin2006,ao2007existence,ao2008cancer,Ao2008,Ao2009,xu2011landscape,Shi2012}
have well been expanded into biology and other fields, generating new understandings on stochastic calculus.
The novel framework \cite{Ao2004,Kwon2005,Yin2006,ao2007existence} in the present brief review
is \textit{beyond} It\^{o}-Stratonovich controversy from two aspects:
\textit{First}, it defines a new type (named \textit{A-type} for short) of stochastic integration that is different from It\^{o} and Stratonovich's. In one dimensional case, A-type integration reduces to the $\alpha$-type \cite{McClintock} with $\alpha=1$ (It\^o's corresponds to $\alpha=0$ and Stratonovich's to $\alpha=0.5$). In higher dimensional situations, it is generally not an $\alpha$-type integration \cite{Shi2012}.
\textit{Second}, there is a starting point in which obtaining this new approach,
we need not to choose the ``correct" calculus. One can reach the same result by It\^o, Stratonovich or A-type prescription of stochastic integration.

The effectiveness of the new framework has been demonstrated by its various applications:
The successful solution of the outstanding stability puzzle of a genetic switch \cite{Zhu2004};
the quantification of evolutionary dynamics of Darwin and Wallace \cite{ao2005laws};
the study of complex bio-networks such as metabolic network \cite{ao2005metabolic} and cancer network \cite{ao2008cancer};
the implications of Darwinian dynamics in physics \cite{Ao2008};
the relationship with Lyapunov's direct method for stability analysis in engineering \cite{yuan2011potential};
the explicit construction of adaptive landscape in population genetics \cite{xu2011landscape} and etc.
Results of a recent experiment in one dimension \cite{volpe2010influence} provide evidence
that there exist processes in Nature choosing A-type integration.

A transformation from the classical Langevin equation to a structured form lies at the core of this framework.
The significance of such transformation is the obtaining of a potential function that plays a dual role:
It leads to the Boltzmann-Gibbs distribution on the final steady state (if it exists) of the stochastic process;
it corresponds to the deterministic dynamics as a global Lyapunov function and can be used for stability analysis \cite{yuan2011potential}.
The potential function exists for general non-equilibrium processes without detailed balance \cite{Ao2004,Yin2006,ao2007existence},
which are usually difficult to handle by previous methods \cite{blythe2007nonequilibrium}.
The behaviors near a fixed point, stable or not, have been exhaustively studied \cite{Kwon2005};
explicit construction for a simple limit cycle dynamics has been provided \cite{Zhu2006} as well.
The structured form itself reveals the symplectic structure embedded in stochastic differential equations and
has an invariant property \cite{ao2007existence} during the transformation.
Other applications based on the concept of potential function have also been studied in biology \cite{Frauenfelder1991,xing2007nonequilibrium,Wang2008,Ao2009,Qian2010,karl2011},
physics \cite{wang2003a,ge2009,xing2010mapping} and control theory \cite{Johansson1998,wang2003}.

In the following, we first review the framework in Sect.~\ref{framework} about the transformation and its consistency in mathematics.
In Sect.~\ref{discussion}, several related issues have been discussed, including a straightforward implication on the physical meaning.
In particular, the one dimensional experiment serving as an example is analyzed in Sect.~\ref{one-dim}.
We summarize in Sect.~\ref{conclusion}.

\section{The framework}\label{framework}

A heuristic demonstration of this framework is in the following subsection.
It is first proposed by one of the authors in \cite{Ao2004}. A more rigorous exposition \cite{Yin2006} is then provided in Sect.~\ref{derivation}.

\subsection{Review of the Transformation}\label{trans}

Stochastic differential equation, or the Langevin equation in physics, is usually a more precise description of reality than the purely deterministic one \cite{Gardiner2004,Kampen2007,risken1996fokker}. Additive noise is frequently considered for physical systems, but for biological or other complex systems, a state variable dependent (multiplicative) noise is often encountered. In this brief review, we use the physicists' notation with a multiplicative noise:
\begin{align}
\label{Langevin}
\dot{\mathbf{q}}=\mathbf{f}(\mathbf{q})+N(\mathbf{q})\xi(t)~,
\end{align}
where $\mathbf{q}$, $\mathbf{f}$ are $n$-dimensional vectors and $\mathbf{f}$ is a nonlinear function of the state variable $\mathbf{q}$. The noise $\xi(t)$ is $k$-dimensional Gaussian white with the zero mean, $\left<\xi(t)\right>=0$, and the covariance $\left<\xi(t)\xi^\tau(t')\right>=\delta(t-t')I_k$. The superscript $\tau$ denotes the transpose of a matrix, $\delta(t-t')$ is the Dirac delta function, $\left<...\right>$ indicates the average over noise distribution, $I_k$ is the $k$-dimensional identity matrix. The element of the $n\times k$ matrix $N(\mathbf{q})$ can be a nonlinear function of $\mathbf{q}$. This matrix is further described by:
\begin{align}
\label{varianceD}
   N(\mathbf{q})N^\tau(\mathbf{q})=2\epsilon D(\mathbf{q})~,
\end{align}
where $\epsilon$ is a constant quantifying the noise strength and $D(\mathbf{q})$ is an $n\times n$ positive semi-definite diffusion matrix.
Note that if the noise has less than $n$ independent components, $k< n$, $D(\mathbf{q})$ has zero eigenvalue(s).

During the study of a biological switch \cite{Zhu2004}, a structured form equivalent to Eq.~\eqref{Langevin} was discovered:
\begin{align}
\label{decomp}
   [S(\mathbf{q})+A(\mathbf{q})]\dot{\mathbf{q}}=-\nabla\phi(\mathbf{q})+\hat{N}(\mathbf{q})\xi(t)~,
\end{align}
where $S(\mathbf{q})$ is symmetric and positive semi-definite, $A(\mathbf{q})$ is antisymmetric, $\phi(\mathbf{q})$ is the potential function, $\xi(t)$ is identical to that in Eq.~\eqref{Langevin} and the matrix $\hat{N}(\mathbf{q})$ is constrained by (see also Sect.~\ref{physical_meaning}):
\begin{align}
\label{variance}
\hat{N}(\mathbf{q})\hat{N}^\tau(\mathbf{q})=2\epsilon S(\mathbf{q})~.
\end{align}
We note that when $S=0$, Eq.~\eqref{decomp} has the same structure as the Hamiltonian equation in physics. The $(n=2l)$-dimensional system ($l$ generalized coordinates and $l$ generalized momentums combine the $n$-dimensional system) with
\begin{align*}
 A =\left(\begin{array}{cc}0 &I_l\\-I_l &0\end{array}\right)
\end{align*}
and $\phi=H$ (the Hamiltonian), shows the embedded symplectic structure.

There are two assumptions implied in this framework:
The \textit{first} one is the equivalence of Eq.~\eqref{Langevin} and Eq.~\eqref{decomp},
that is, they are describing the same dynamical process.
From Eq.~\eqref{decomp} to Eq.~\eqref{Langevin} is straightforward, once $[S(\mathbf{q})+A(\mathbf{q})]$ is nonsingular,
which holds when the components $q_i(i=1,\dots,n)$ of the state variable $\mathbf{q}=(q_1,\dots,q_n)^\tau$ are independent.
The reverse, however, is much more difficult, including to obtain $S(\mathbf{q})$, $A(\mathbf{q})$ and $\phi(\mathbf{q})$ for general dynamics.
Nevertheless, we can still assume, without a rigorous mathematical proof but rigorous enough from a working scientist's view,
the reverse part is held based on a case by case construction, that is, considering it as a protocol not a theorem.
Hence, we may replace $\dot{\mathbf{q}}$ with the right hand side of Eq.~\eqref{Langevin} in Eq.~\eqref{decomp}:
\begin{align}
\label{replace}
  \left[ S(\mathbf{q})+A(\mathbf{q}) \right]\left[\mathbf{f}(\mathbf{q})+N(\mathbf{q})\xi(t)\right]=-\nabla \phi (\mathbf{q})+\hat{N}(\mathbf{q})\xi(t)~.
\end{align}

The \textit{second} assumption is the separated equality of the deterministic and stochastic dynamics in Eq.~\eqref{replace}:
\begin{align}
\label{determin}
   \left[ S(\mathbf{q})+A(\mathbf{q}) \right]\mathbf{f}(\mathbf{q})=-\nabla \phi (\mathbf{q})~,\\
\label{stochastic}
   \left[ S(\mathbf{q})+A(\mathbf{q}) \right]N(\mathbf{q})=\hat{N}(\mathbf{q})~.
\end{align}
Intuitively, this assumption on separation is plausible for two different reasons:
\textit{First}, the zero mean noise function is nowhere differentiable
but the deterministic forces are usually smooth ($C^{\infty}$) functions, hence two very different mathematical objects;
\textit{second}, the stochastic force and the deterministic forces describe different timescales of a physical system
with different physical origins.
By replacing Eq.~\eqref{determin} with an equivalent form \eqref{curl} and plugging Eq.~\eqref{stochastic} and Eq.~\eqref{varianceD} into Eq.~\eqref{variance},
we obtain the potential condition Eq.~\eqref{curl} and the generalized Einstein relation Eq.~\eqref{GER}:
\begin{align}
\label{curl}
  &\nabla\times\left\{\left[ S(\mathbf{q})+A(\mathbf{q})\right]\mathbf{f}(\mathbf{q})\right\}=0~, \\
\label{GER}
 [S(\mathbf{q})&+A(\mathbf{q})]D(\mathbf{q})[S(\mathbf{q})-A(\mathbf{q})]=S(\mathbf{q})~.
\end{align}
In principle, the potential function $\phi(\mathbf{q})$ can be derived analytically by solving the $n(n-1)/2$ partial differential equations
(under proper boundary conditions) Eq.~\eqref{curl}, together with the $n(n+1)/2$ equations given
by Eq.~\eqref{GER} ($n^2$ unknowns in $[S(\mathbf{q})+A(\mathbf{q})]$ and $n^2$ equations).
It can also be calculated numerically through a gradient expansion \cite{Ao2004}. Detailed results near fixed points can be found in \cite{Kwon2005}. Explicit constructions of potential functions for limit cycles are contained in \cite{Zhu2006,yuan2011potential}.

In one dimensional case, $A = 0$, let $\epsilon=k_BT$, if the friction $\gamma$ is a constant, then $S=\gamma/k_BT$, Eq.~\eqref{GER} reduces to $SD=\gamma D/k_BT=1$, namely, the product of the friction and diffusion coefficients is a constant, discovered by Einstein \cite{Einstein1905} one hundred years ago.
Equation \eqref{GER} is a generalized form of the Einstein relation in two ways:
The diffusion matrix can be nonlinear dependent of the state variable (a verification experiment has been reported \cite{prieve1999measurement})
and the detailed balance condition can be broken ($A(\mathbf{q})\neq 0$, see also Sect.~\ref{zero_mass}).

From the generalized Einstein relation Eq.~\eqref{GER}, we know the symmetric part of $[S(\mathbf{q})+A(\mathbf{q})]^{-1}$, $D(\mathbf{q})=1/2\left[(S(\mathbf{q})+A(\mathbf{q}))^{-1}+((S(\mathbf{q})+A(\mathbf{q}))^{-1})^\tau\right]$, is the diffusion matrix defined in Eq.~\eqref{varianceD}. Hence we have $[S(\mathbf{q})+A(\mathbf{q})][D(\mathbf{q})+Q(\mathbf{q})]=I$ ($I$ is the identity matrix, singularity of $D(\mathbf{q})$, $S(\mathbf{q})$ or $A(\mathbf{q})$ can usually be tolerated) where $Q(\mathbf{q})$ is antisymmetric. Rewrite Eq.~\eqref{determin} as
\begin{align}
\label{standard}
\mathbf{f}(\mathbf{q})=-\left[ D(\mathbf{q})+Q(\mathbf{q}) \right]\nabla \phi (\mathbf{q})~,
\end{align}
and by multiplying $\nabla \phi (\mathbf{q})$ on both sides of Eq.~\eqref{standard}, we reach the Hamilton-Jacobi equation:
\begin{align}
\label{HJ}
\mathbf{f}(\mathbf{q})\cdot \nabla \phi (\mathbf{q})+\nabla \phi (\mathbf{q})\cdot D(\mathbf{q})\cdot \nabla \phi (\mathbf{q})=0~.
\end{align}

Finally, by the knowledge from statistical physics, we can consider the potential function leads to a Boltzmann-Gibbs distribution
\begin{align}
\label{BG}
   \rho_s(\mathbf{q}) = \frac{1}{Z(\epsilon)}\exp\left(-\frac{\phi(\mathbf{q})}{\epsilon}\right)~,
\end{align}
as the steady state distribution (if it exists) for the stochastic process described by Eq.~\eqref{decomp}:
 $\epsilon$ as the ``temperature" and $\phi(\mathbf{q})$ as the ``Hamiltonian".
Experimental verification is available for some cases, e.g. \cite{volpe2010influence}.

Hence, the two assumptions to implement the equivalence between Eqs.~\eqref{Langevin} and \eqref{decomp}
lead to a potential function for general Langevin equation and to a corresponding Boltzmann-Gibbs distribution for the final steady state independent of both matrices $A(\mathbf{q})$ and $S(\mathbf{q})$. While such distribution function seems special, it should be pointed out that it is one of the most successful equations in physics,
even broader than that of the stationary Schr\"odinger equation.
The framework can be applied for systems with an arbitrary noise strength:
Eqs.~\eqref{curl} and \eqref{GER} do not contain $ \epsilon$.
Particularly, deterministic systems can be regarded as the weak noise limit ($\epsilon\to0$) of stochastic systems.
The framework is also able to be extended directly to include parameter dependent potential
and corresponding dynamical components \cite{Zhu2006,Ao2008}, e.g. time $t$ and temperature $\epsilon$ dependent $\phi(\mathbf{q},t,\epsilon)$.
We note that the Eqs.~\eqref{determin} and \eqref{GER} have a close connection to Lyapunov's direct method in engineering \cite{yuan2011potential}.
These evidences make the assumptions more plausible and appealing.

This construction further leads to a stochastic integration (A-type) different from conventional ones such as It\^{o} or Stratonovich.
One would ask the mathematical consistency in this new approach.
We point out that the consistency of A-type integration may be seen from two aspects:
 \textit{First}, Eq.~\eqref{Langevin} or \eqref{decomp} by themselves are not complete descriptions of a stochastic process.
A proper stochastic interpretation is needed, hence A-type is possible in principle.
 \textit{Second}, we will show in the next subsection that there is an important situation in which one starts without
the need to differentiate various known stochastic integrations,
and reaches the consequence that A-type interpretation is the natural choice for the SDE Eqs.~\eqref{decomp} and \eqref{Langevin}.

\subsection{Demonstration of Consistency}\label{derivation}
In this subsection, we summarize the main steps to demonstrate that A-type integration is mathematically consistent.
The rationality is as follows:
The starting point is the important stochastic differential equation in physics, Newton equation in $2n$ dimensions with noise and friction.
Its corresponding $2n$-dimensional Fokker-Planck equation (FPE) is uniquely determined, known as generalized Klein-Kramers equation.
The valuable feature here is, all approaches to stochastic integration discussed in the present brief review lead to the same $2n$-dimensional FPE.
When taking the zero mass limit, the $2n$ dimensional Newton equation reduces to an $n$ dimensional equation exactly in the same form as Eq.~\eqref{decomp},
while the corresponding $n$ dimensional FPE derived from the generalized Klein-Kramers equation has the ability
to permit the Boltzmann-Gibbs distribution as its stationary solution.
After this zero-mass limit, the connection between the resulting $n$ dimensional FPE
and the stochastic differential equation is the A-type integration.

\subsubsection{Generalized Klein-Kramers Equation}
We begin with a general situation via doubling degrees of freedom, the $2n$-dimensional equation
with noise, proposed in \cite{Ao2004}:
\begin{equation}
\label{2nform}
\left\{
\begin{array}{l}
 \dot{\mathbf{q}}=\frac{\mathbf{p}}{m}\\
 \dot{\mathbf{p}}=-[S(\mathbf{q})+A(\mathbf{q})]\frac{\mathbf{p}}{m}-\nabla_\mathbf{q}\phi(\mathbf{q})+\hat{N}(\mathbf{q})\xi(t)
\end{array}
\right.
\end{equation}
where $S(\mathbf{q})$, $A(\mathbf{q})$, $\phi(\mathbf{q})$, $\hat{N}(\mathbf{q})$ and $\xi(t)$ are identical to those
in Eq.~\eqref{decomp}, the subscript of $\nabla_\mathbf{q}$ means it operates on $\mathbf{q}$ only, the same as in Eq.~\eqref{decomp}.
We note that Eq.~\eqref{2nform} has two properties:
\textit{First}, by taking the zero mass limit $m\to 0$, $\mathbf{p}\to 0$, we can recover Eq.~\eqref{decomp} from Eq.~\eqref{2nform};
\textit{Second}, we wish to emphasize that starting with Eq.~\eqref{2nform},
we can find that different stochastic interpretations It\^{o}, Stratonovich, or A-type, will reach the same result,
since $\nabla_\mathbf{p}\cdot\left[S(\mathbf{q})+A(\mathbf{q})\right]=0$.

The probability density function in the $(\mathbf{q}, \mathbf{p})$ phase space can be calculated through path integral as:
\begin{align}
\label{density}
\rho(\mathbf{q}, \mathbf{p}, t)\equiv\left<\delta\left (\mathbf{q}-\bar{\mathbf{q}}\left(t,\{\xi(t)\}\right)\right) \delta\left(\mathbf{p}-\bar{\mathbf{p}}\left(t,\{\xi(t)\}\right)\right)\right>~,
\end{align}
where $\bar{\mathbf{q}}\left(t,\{\xi(t)\}\right)$ and $\bar{\mathbf{p}}\left(t,\{\xi(t)\}\right)$ are the solution of the ordinary differential equation (ODE) reduced from Eq.~\eqref{2nform} under a fixed sample path $\{\xi(t)\}$ of the sample space of $\xi(t)$ (a given noise configuration). By taking partial derivative of time on both sides of Eq.~\eqref{density}, we obtain
\begin{align}
\partial_t\rho&=\left<\nabla_\mathbf{\bar{\mathbf{q}}}\delta\left(\mathbf{q}-\bar{\mathbf{q}}\right)
\cdot
\frac{d\bar{\mathbf{q}}}{dt}
\delta\left(\mathbf{p}-\bar{\mathbf{p}}\right)+
\delta\left(\mathbf{q}-\bar{\mathbf{q}}\right)
\nabla_\mathbf{\bar{\mathbf{p}}}\delta\left(\mathbf{p}-\bar{\mathbf{p}}\right)
\cdot
\frac{d\bar{\mathbf{p}}}{dt}\right>\nonumber\\
&=-\int\delta\left(\mathbf{q}-\bar{\mathbf{q}}\right)\delta\left(\mathbf{p}-\bar{\mathbf{p}}\right)\left\{ \nabla_{\bar{\mathbf{q}}} \cdot \left[\dot{\bar{\mathbf{q}}} P(\{\xi\})\right]+\nabla_{\bar{\mathbf{p}}} \cdot \left[\dot{\bar{\mathbf{p}}} P(\{\xi\})\right] \right\}d\{\xi\}\nonumber\\
&=-\nabla_\mathbf{q}\cdot\frac{\mathbf{p}}{m}\rho-\nabla_\mathbf{p}\cdot\left\{-\left[S(\mathbf{q})+A(\mathbf{q})\right]\cdot\frac{\mathbf{p}}{m}\rho-\nabla_{\mathbf{q}}\phi(\mathbf{q})\rho\right.\nonumber\\
\label{steps}
&\quad\left.+\hat{N}(\mathbf{q})\left<\xi(t)\delta\left(\mathbf{q}-\bar{\mathbf{q}}\right)\delta\left(\mathbf{p}-\bar{\mathbf{p}}\right)\right>\right\}~,
\end{align}
in which $\rho$, $\bar{\mathbf{q}}$ and $\bar{\mathbf{p}}$ have the same meaning as in Eq.~\eqref{density}, $P(\{\xi\})$ is the probability density function for the given noise configuration $\{\xi(t)\}$. From step 1 to step 2 of Eq.~\eqref{steps}, integration by parts is used.
According to the Novikov identity \cite{novikov1965functionals}, for a given functional $f[\xi(t),t]$ of noise $\xi(t)$,
\begin{align}
\left<\xi(t)f[\xi(t),t]\right>&=
\int\left<\xi(t)\xi^\tau(t')\right>\left<\frac{\delta f[\xi(t),t]}{\delta \xi(t')}\right>dt'\nonumber\\
&=\left<\frac{\delta f[\xi(t),t]}{\delta \xi(t)}\right>~,
\end{align}
and using
\begin{align}
\frac{\delta \left[\int_{0}^t\xi(t')dt'\right]}{\delta \xi(t)}=\frac{1}{2}~,
\end{align}
noting that the solution of Eq.~\eqref{2nform} can be expressed as
\begin{align}
\bar{\mathbf{q}}(t)&=\mathbf{q}(0)+ \int_{0}^t \frac{\bar{\mathbf{p}}}{m}dt'~,\\
\bar{\mathbf{p}}(t)&=\mathbf{p}(0)+ \int_{0}^t \left\{-\left[S(\bar{\mathbf{q}})+A(\bar{\mathbf{q}})\right]\frac{\bar{\mathbf{p}}}{m}-\nabla_{\bar{\mathbf{q}}} \phi(\bar{\mathbf{q}})+ \hat{N}(\bar{\mathbf{q}})\xi(t')\right\}dt'~,
\end{align}
then we obtain the relations:
\begin{align}
\frac{\delta\bar{\mathbf{q}}(t)}{\delta\xi(t)}&=0~,\\
\frac{\delta\bar{\mathbf{p}}(t)}{\delta\xi(t)}&=\frac{1}{2}\hat{N}^\tau(\bar{\mathbf{q}})~.
\end{align}
The last left term in Eq.~\eqref{steps} can be handled as
\begin{align}
\hat{N}(\mathbf{q})\left<\xi(t)\delta\left(\mathbf{q}-\bar{\mathbf{q}}\right)\delta\left(\mathbf{p}-\bar{\mathbf{p}}\right)\right>= -2\epsilon S(\mathbf{q})\cdot \frac{1}{2}\nabla_\mathbf{p}\rho~.
\end{align}
We finally reach the generalized version of the Klein-Kramers equation \cite{Kampen2007}:
\begin{align}
\label{GKK}
\partial_t\rho=\nabla_\mathbf{p}\cdot\left\{\left[A(\mathbf{q})+S(\mathbf{q})\right]\cdot
\frac{\mathbf{p}}{m}+\nabla_{\mathbf{q}}\phi(\mathbf{q})+\epsilon  S(\mathbf{q})\cdot \nabla_{\mathbf{p}}\right\}\rho-\frac{\mathbf{p}}{m}\cdot\nabla_\mathbf{q} \rho~.
\end{align}
Note that Eq.~\eqref{GKK} has a solution (if the partition function $Z(\epsilon)$ is finite)
\begin{align}
\label{2nsteady}
\rho_s(\mathbf{q},\mathbf{p})=\frac{1}{Z(\epsilon)}\exp\left(-\frac{\frac{p^2}{2m}+\phi(\mathbf{q})}{\epsilon}\right)~.
\end{align}
Equation \eqref{2nsteady} is not explicitly time dependent, $\partial_t\rho_s=0$, it is a steady state distribution.
The state variables $\mathbf{q}$ and $\mathbf{p}$ are separated, which simplifies the process of taking zero mass limit to be discussed below.

\subsubsection{Zero Mass Limit}\label{zero_mass}

We have mentioned that after taking zero mass limit $m\to0$, Eq.~\eqref{2nform} reduces to Eq.~\eqref{decomp}.
Therefore, by acting the same limit process on Eq.~\eqref{GKK}, the corresponding FPE to Eq.~\eqref{decomp} can be obtained \cite{Yin2006}:
\begin{align}
\label{FPE}
  \partial_t \rho(\mathbf{q},t)=\nabla\cdot\left[D(\mathbf{q})+Q(\mathbf{q})\right]\cdot[\epsilon\nabla+\nabla\phi(\mathbf{q})]
    \rho(\mathbf{q},t)~,
\end{align}
where $D(\mathbf{q})$ is the diffusion matrix, $Q(\mathbf{q})$ is defined in Eq.~\eqref{standard}, $\nabla\equiv\nabla_\mathbf{q}$, and the derivative $\nabla$ in $\nabla\phi(\mathbf{q})$ does not operate on $\rho(\mathbf{q},t)$.
This FPE has a drift velocity $-\left[D(\mathbf{q})+Q(\mathbf{q})\right]\nabla\phi(\mathbf{q})=\mathbf{f}(\mathbf{q})$
and a generalized symmetric form $\nabla\cdot\left[D(\mathbf{q})+Q(\mathbf{q})\right]\cdot\epsilon\nabla\rho(\mathbf{q},t)$.
The commonly defined probability current density $\mathbf{j}(\mathbf{q},t)=(j_1(\mathbf{q},t),\dots,j_n(\mathbf{q},t))^\tau$ is:
\begin{align}
  j_i(\mathbf{q},t)=(f_i+\Delta f_i)(\mathbf{q})\rho(\mathbf{q},t)-\partial_{j} \left[\epsilon D_{ij}(\mathbf{q})\rho(\mathbf{q},t)\right]
\end{align}
where $\mathbf{f}(\mathbf{q})=(f_1(\mathbf{q}),\dots,f_n(\mathbf{q}))^\tau$,
$\Delta f_i(\mathbf{q})=\epsilon\partial_j[D_{ij}(\mathbf{q})+Q_{ij}(\mathbf{q})]$ (see also Sect.~\ref{relation}),
 with $i=1,\dots,n$ and $j=1,\dots,n$, $D_{ij}(\mathbf{q})$ the $i$th-row and $j$th-column element of the diffusion matrix $D(\mathbf{q})$, $Q_{ij}(\mathbf{q})$ the element of
 $Q(\mathbf{q})$ in Eq.~\eqref{FPE}.
 For steady state, Eq.~\eqref{BG}, $\nabla\cdot\mathbf{j}_s(\mathbf{q})=0$.
One can calculate that when $Q=0$ then $\mathbf{j}_s=0$;
if $Q(\mathbf{q})\neq 0$ then generally $\mathbf{j}_s(\mathbf{q})\neq0$,
since $\partial_j\left[Q_{ij}(\mathbf{q})\rho_s(\mathbf{q})\right]\neq0$. Therefore the framework encompasses the cases without detailed balance.
The term ``detailed balance" used here is defined for abstract phase space,
that is, the net current between any two states in the phase space is zero \cite{kubo1995},
identical to that for Markov process in mathematics.

There are various ways to take the zero mass limit (eliminate the fast degrees $\mathbf{p}$),
a derivation adopting the standard projection operator method \cite{Gardiner2004} is given in \cite{Yin2006},
that is, to project the $(\mathbf{q},\mathbf{p})$ space to the $\mathbf{q}$ space.
We note that another commonly used approach of degree reduction, adiabatic elimination \cite{sancho1982,Gardiner2004,McClintock},
is based on the postulation that the velocity is a fast variable, which is consistent with the zero mass limit.
Three other derivations for 1-dimensional systems are discussed in \cite{ao2007existence}.
This FPE (Eq.~\eqref{FPE}) defines the A-type integration for SDE of the form of Eq.~\eqref{decomp}.
Since generally Eq.~\eqref{Langevin} is equivalent to Eq.~\eqref{decomp},
Eq.~\eqref{FPE} defines a stochastic integration for general Langevin equations.
The steady state distribution in the phase space (if it exists) is the Boltzmann-Gibbs distribution Eq.~\eqref{BG}.
We should clarify that it is not to say that A-type is the only way to integrate Eq.~\eqref{decomp},
but a stochastic interpretation with physical significance.

\section{Discussions}\label{discussion}

\subsection{Physical Meaning of Equation \eqref{decomp}}\label{physical_meaning}

We can consider a time-evolution system described by SDE as a particle motion inside the phase space. From a physical point of view, the motion of an object should be driven by some underlying forces. We consider this particle as a charged and \textit{massless} one driven by the force:
$\mathbf{F}_{total}=m\ddot{\mathbf{q}}=0$  ($m=0$).
The driven force can be separated generally in physics into three parts, dissipative, conservative and random:
\begin{align}
\mathbf{F}_{total}=\mathbf{F}_{dissipative}+\mathbf{F}_{conservative}+\mathbf{F}_{random}=0~.
\end{align}
Equation \eqref{decomp} has exactly this structure. In the usual study of 2 and 3 dimensional cases, Eq.~\eqref{decomp} corresponds to a known fundamental equation in physics.
We can generally use a frictional force as the dissipative component
$\mathbf{F}_{dissipative}=-S(\mathbf{q})\dot{\mathbf{q}}$ where $S(\mathbf{q})$ is symmetric and positive semi-definite; a Lorentz force together with an energy induced force (for example, the electrostatic force) as the conservative component $\mathbf{F}_{conservative}=e\dot{\mathbf{q}}\times \mathbf{B}(\mathbf{q})+\left[-\nabla \phi(\mathbf{q})\right]$; the random component is described by a Gaussian white noise $\mathbf{F}_{random}=\hat{N}(\mathbf{q})\xi(t)$ with zero mean that has the common origin with the frictional force, as formulated by the fluctuation dissipation theorem \cite{callen1952theorem,callen1951irreversibility,Kubo1966}: $\hat{N}(\mathbf{q})\hat{N}^\tau(\mathbf{q})=2\epsilon S(\mathbf{q})$, hence
\begin{align}
\label{temp}
-S(\mathbf{q}) \dot{\mathbf{q}}+e\dot{\mathbf{q}}\times \mathbf{B}(\mathbf{q})-\nabla \phi(\mathbf{q})+\hat{N}(\mathbf{q})\xi(t)=0~.
\end{align}
The friction matrix $S(\mathbf{q})$ is positive semi-definite (guaranteed by the fluctuation-dissipation theorem), keeping the resistant property of the dissipative force (the allowed values are confined in the negative half space). It can be a non-diagonal matrix, describing an anisotropic frictional force. The potential function $\phi(\mathbf{q})$ plays the similar role to the Hamiltonian in dissipative systems, leading to the Boltzmann-Gibbs distribution on the final steady state (if it exists) of the stochastic process.
For higher dimension, we generalize $\mathbf{B}(\mathbf{q})\times\dot{\mathbf{q}}$ as $A(\mathbf{q})\dot{\mathbf{q}}$,
where $A(\mathbf{q})$ is an antisymmetric matrix.
There is a direct and simple physical realization based on this intuitive explanation of the SDE Eq.~\eqref{decomp},
for instance, practical experiments to verify the generalized Einstein relation are designed in Section 5.3 of \cite{Ao2008}.
A recent experiment \cite{volpe2010influence} has been implemented under a similar physical setting.

\subsection{Difference and Relation of the A-type Integration with Traditional Ones}\label{relation}

We have mentioned that the A-type integration reduces to an $\alpha$-type one with $\alpha=1$ for one dimensional system, hence it is different from It\^o or Stratonovich's. A more detailed demonstration on this point is provided in \cite{ao2007existence}. For general cases, when dimension is higher than one, things become more complicated, A-type integration is even not $\alpha$-type, that means at different time intervals the choices of $\alpha$ are not the same, or the parameter $\alpha$ is position dependent $\alpha(\mathbf{q})$. A relation between A-type integration and It\^o's is formulated in \cite{Shi2012} by considering the correspondence of their FPEs, that is: the SDE
\begin{align}
\label{sys1}
\dot{\mathbf{q}}=(\mathbf{f}+\Delta \mathbf{f})(\mathbf{q})+N(\mathbf{q})\xi(t)~,
\end{align}
where $\Delta \mathbf{f}(\mathbf{q})=(\Delta f_1(\mathbf{q}),\dots,\Delta f_n(\mathbf{q}))^\tau$ using It\^{o} integration describes the same process to that of the SDE
\begin{align}
\label{sys2}
\dot{\mathbf{q}}=\mathbf{f}(\mathbf{q})+N(\mathbf{q})\xi(t)~,
\end{align}
using A-type integration. Here $\Delta f_i(\mathbf{q})=\epsilon\partial_j[D_{ij}(\mathbf{q})+Q_{ij}(\mathbf{q})]$.
We note that there is an obvious distinction between $\Delta \mathbf{f}(\mathbf{q})$ here and the additional (to It\^o's) drift term for $\alpha$-type ($\alpha$ choosing as a constant) integrations $\Delta f_i^{(\alpha)}(\mathbf{q})=\alpha \left[\partial_k N_{ij}(\mathbf{q})\right]N_{kj}(\mathbf{q})$.

An illustrative example has been provided in \cite{Shi2012} to show a significant difference on the final steady state distributions between It\^o and A-type integrations, the dynamics $\dot{\mathbf{q}}=\mathbf{f}(\mathbf{q})+N(\mathbf{q})\xi(t)$ is \footnote{Note that in the published version in \textit{J. Stat. Mech.}, there is a typo with the noise term in Eq.~(31) and Eq.~(33): a coefficient $\sqrt{2}$ is missing. The $\epsilon$ should be 1.}
\begin{align}
\label{first}
\left\{\begin{array}{rcr}
\dot{x}&=&2x-x\left(x^2+y^2\right)+\sqrt{2(x^2+y^2)}\xi_x(t)\\
\dot{y}&=&2y-y\left(x^2+y^2\right)+\sqrt{2(x^2+y^2)}\xi_y(t)
\end{array}\right.
\end{align}
with $\mathbf{q}=(x,y)^\tau$, $\epsilon=1$, the diffusion matrix $D(x,y)=\left(x^2+y^2\right)\cdot I_2$, $\xi(t)=(\xi_x,\xi_y)^\tau$ has zero mean, and $\left<\xi(t)\xi^\tau(t')\right>=\delta(t-t')I_2$. The final steady state distribution of this system with A-type integration can be calculated based on the framework as $\rho_{sA}(x,y)=\frac{1}{Z_A(\epsilon)}\exp\left[{-\phi/ \epsilon}\right]$ with
\begin{align}
\label{phi}
\phi(\mathbf{q})=-ln\left(x^2+y^2\right)+\frac{x^2+y^2}{2}~.
\end{align}
The distribution of It\^o integration for the system Eq.~\eqref{first} is identical to the A-type's distribution for the system:
\begin{align}
\left\{\begin{array}{rcr}
\dot{x}&=&-x\left(x^2+y^2\right)+\sqrt{2(x^2+y^2)}\xi_x(t)\\
\dot{y}&=&-y\left(x^2+y^2\right)+\sqrt{2(x^2+y^2)}\xi_y(t)
\end{array}\right.
\end{align}
whose expression can be similarly calculated as $\rho_{sI}(x,y)=\frac{1}{Z_I(\epsilon)}\exp\left[{-\psi/\epsilon}\right]$ with
\begin{align}
\label{psi}
\psi(\mathbf{q})=\frac{1}{2}\left(x^2+y^2\right)~.
\end{align}
Similarly, It\^o integration for the system
\begin{align}
\left\{\begin{array}{rcr}
\dot{x}&=&4x-x\left(x^2+y^2\right)+\sqrt{2(x^2+y^2)}\xi_x(t)\\
\dot{y}&=&4y-y\left(x^2+y^2\right)+\sqrt{2(x^2+y^2)}\xi_y(t)
\end{array}\right.
\end{align}
will reach the distribution $\rho_{sA}(x,y)$.

\begin{figure}
\begin{center}
  \subfigure[It\^o Integration]{\includegraphics[width=0.4\textwidth]{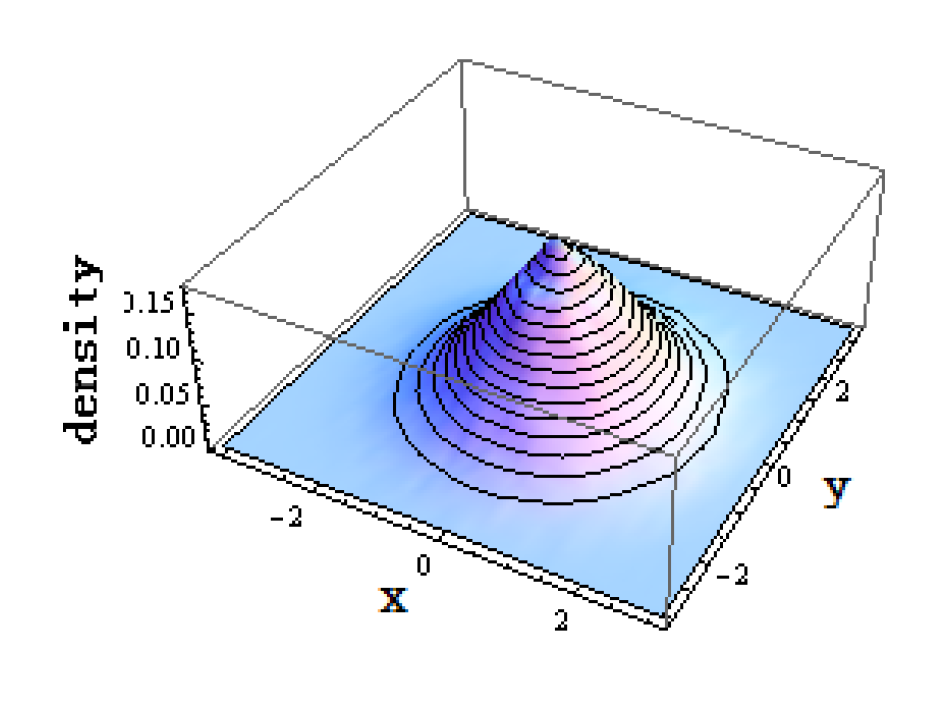}} \quad
  \subfigure[A-type Integration]{\includegraphics[width=0.4\textwidth]{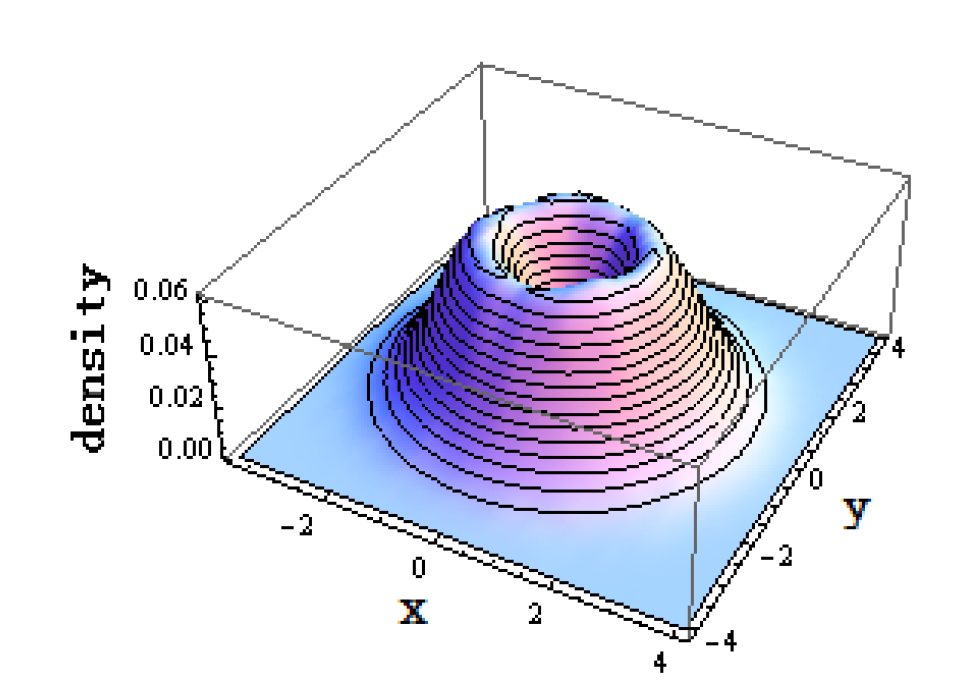}}
\end{center}
\caption{Sampling Results: The figure shows the distributions in the phase space for the long-term sampling of Eq.~\eqref{first} using It\^o and A-type integrations, the basal plane denotes the two dimensional phase space, the vertical axis indicates the emerging probability density of the system at a specific state. Fig.~(a) is for It\^o integration, it is consistent with the theoretical result $\rho_{sI}(x,y)\propto\exp\left[{-\psi/\epsilon}\right]$ ($\psi$ is given in Eq.~\eqref{psi}, with $(0,0)$ being the most probable state). Fig.~(b) is for A-type integration, it is coherent with the deterministic dynamics obtained when $\epsilon=0$, for example, being most probable at the circle $x^2+y^2=2$ (are stable fixed points when $\epsilon=0$), and having zero probability at $(0,0)$ (is an unstable fixed point when $\epsilon=0$).}
\label{fig}
\end{figure}

The two distributions $\rho_{sA}$ and $\rho_{sI}$ have obvious differences, for instance, $\rho_{sA}(0,0)=0$ but the origin $(0,0)$ is the most probable state for $\rho_{sI}$. These theoretical results have been verified by numerical experiments in \cite{Shi2012} (shown in Fig.~\ref{fig}).
Note that the potential $\phi(\mathbf{q})$ in Eq.~\eqref{phi} serves as a global Lyapunov function \cite{yuan2011potential} for the deterministic dynamics (when $\epsilon=0$) of the system Eq.~\eqref{first}. The Lie derivative \cite{yano1957theory} of $\phi(\mathbf{q})$ ($\nabla\phi(\mathbf{q})\cdot \mathbf{f}(\mathbf{q})$) keeps non-positive. Therefore, there is an exact correspondence between the deterministic dynamics and the steady state distribution derived using A-type integration for SDEs, for example, the stable fixed points are locally most probable states, in accordance with one's intuition. This direct correspondence can not be kept by applying It\^o nor Stratonovich integration: After using It\^o's, the unstable point $(0,0)$ of Eq.~\eqref{first} ($\epsilon=0$) becomes stable in $\psi(\mathbf{q})$; moreover, stable points on $x^2+y^2=2$ in Eq.~\eqref{first} ($\epsilon=0$) disappear in Eq.~\eqref{psi}. The unique advantage of using A-type integration comparing to It\^o or Stratonovich's is shown here. It enables a straightforward calculation of the transition probability from one fixed point to another after taking into account the noise influence for ODE models which is essential in many applications \cite{Zhu2004,ao2005laws,ao2005metabolic,ao2008cancer,Ao2009}. For It\^o or Stratonovich integrations, this is not direct and can even be impossible, since one can not recognize the original stable fixed points from the long time sampling distribution, like what shows in this example.

\subsection{1-dimensional Example}\label{one-dim}

A comprehensive discussion on 1-dimensional systems has been conducted previously \cite{ao2007existence}. The corresponding FPEs and the steady state distributions for three different type of stochastic integrations (contained in \cite{ao2007existence}) are listed in Table.~\ref{table}.
As has been shown in \cite{smythe1983observation}, there exist real processes choosing Stratonovich integration.
A recent experiment \cite{volpe2010influence} on a one-dimensional physical process suggested that A-type integration of the corresponding SDE is consistent with the experimental data. We show in the following that the zero mass limit is well established based on the physical setting of the experiment.

For the colloidal particle studied in \cite{volpe2010influence},
$\rho=1510~kg/m^3$, $R=655\times10^{-9}~m$, then the volume $V=4/3 \pi R^3=4/3\times\pi\times (655\times10^{-9})^3~m^3= 1.18\times10^{-18}~m^3$, the mass of the particle is
$m=\rho V=1510 \times 1.18\times10^{-18}~kg= 1.78\times10^{-15}~kg$.
The symbol $dt$ denotes the sampling time interval in \cite{volpe2010influence}, within $dt\leq 10~ms$, the authors claim that ``the force acting on the particle can be treated as locally constant." In different experiments, they choose $dt$ around the magnitude of $1~ms$.

We use the simplified formula for the friction coefficient from Mannella \textit{et al.}'s discussion of the same experiment \cite{McClintock}, with $\eta=8.5~mPa~s$ and $z_0=700~nm$:
\begin{align}
\gamma(z) = 6 \pi \eta R \frac{  (z+z_0)}{z}~.
\end{align}
For $z\to\infty$, we have $\gamma_\infty= 6\pi\eta R= 6 \pi\times8.5\times10^{-3}\times655\times10^{-9}~N s/m= 1.1\times10^{-7} N s/m$, then the time needed for the particle to return to equilibrium after acted by
an external force $F(z)$ is
$t(z)= [F(z) / \gamma(z)] / [F(z) / m] = m / \gamma(z)$, $t_\infty=m/ \gamma_\infty
= 1.78\times10^{-15} / 1.1\times10^{-7}~s = 1.6\times10^{-8}~s << 1\times10^{-3}~s=1~ms\approx dt$.
In Fig.~2 of \cite{volpe2010influence}, for instance, at $z=200~nm$, $\gamma(z) = 1.1\times10^{-7}~N s/m \times (200+700)/200 = 5.0\times10^{-7}~N s/m$.
Since $\gamma(z)$ is always larger than $\gamma_\infty$, such that $t(z)<t_\infty<< dt$. We conclude that during the sampling time interval the particle
has been in equilibrium, hence the mass can be considered as zero in
the experiment, realizing the zero mass limit.

\begin{table}[!ht]
\caption{Fokker-Planck equations and steady state  distributions with $\epsilon=1$}
\begin{tabular}{ccc}
\toprule
\textbf{$\alpha$} & \textbf{Fokker-Plank equation}& \textbf{Steady state distribution}\\ \colrule
 $0$ & $\frac{\partial}{\partial t}\rho(q,t)=\left[\frac{\partial}{\partial q}\frac{\partial}{\partial q}D(q)+\frac{\partial}{\partial q}D(q)\phi(q)\right]\rho(q,t)$ & $\frac{1}{Z_I}\left\{\frac{1}{D(q)}\exp\left[-\phi(q)\right]\right\}$\\

 $\frac{1}{2}$ & $\frac{\partial}{\partial t}\rho(q,t)=\left[\frac{\partial}{\partial q}D^{\frac{1}{2}}(q)\frac{\partial}{\partial q}D^{\frac{1}{2}}(q)+\frac{\partial}{\partial q}D(q)\phi(q)\right]\rho(q,t)$ & $\frac{1}{Z_S}\left\{\frac{1}{D^{\frac{1}{2}}(q)}\exp\left[-\phi(q)\right]\right\}$ \\

 $1$ & $\frac{\partial}{\partial t}\rho(q,t)=\left[\frac{\partial}{\partial q}D(q)\frac{\partial}{\partial q}+\frac{\partial}{\partial q}D(q)\phi(q) \right]\rho(q,t)$ & $\frac{1}{Z_A}\exp\left[-\phi(q)\right]$\\
\botrule
\end{tabular}
\label{table}
\end{table}

\subsection{Previous Attempts}

Some related previous attempts have been discussed in our former works \cite{Ao2004,Kwon2005,Ao2008}.
In this subsection we provide a brief summary.
No effort is made to have a complete list, while we do hope to have picked some of the major ones.

To our knowledge, the one dimensional case of the present framework, $\alpha=1$ type integration,
has been mathematically discussed by Wong and Zakai in the 1960s \cite{wong1965convergence}
(see also \cite{Kwon2005,Ao2008}).
Their work provided a method for $\alpha \in [0,1]$.

From a physics perspective Grabert \textit{et al.} \cite{PhysRevA.21.2136} provide a method for nonlinear irreversible processes,
and was discussed by Graham in a long review which we had commented \cite{Ao2004}.
Their method has a restriction based on the singularity of diffusion matrix $D(\mathbf{q})$ (see Eqs.~(2.14-16) in \cite{PhysRevA.21.2136})
and handles the phase variables separately according to a reversibility defined by such singularity.
By contrast, our framework can be used directly for a given $D(\mathbf{q})$ in its general form without a separated treatment,
a global potential is then obtained in the whole phase space.

Theoretical considerations for the situations with detailed balance have been proposed by Klimontovich \cite{klimontovich1990ito} (see also \cite{Ao2008}),
while our approach applies to situations of both with and without detailed balance.
On the weak noise limit, previous works reach a same Hamilton-Jacobi equation to our Eq.~\eqref{HJ}, the nondifferentiable potentials reported in \cite{PhysRevA.31.1109,jauslin1987nondifferentiable} are then included within our framework; moreover, our potential function is valid for an arbitrary noise strength.
A major difference between our construction of the potential and those in the literature such as Graham-Haken construction \cite{Haken} is
that no assumption on the stationary distribution function is needed in our approach in the limit $t\to\infty$.
In particular, the potential in our framework can be time dependent (see also \cite{Ao2004,Zhu2006}).
Other results beyond It\^{o} and Stratonovich have been proposed in \cite{PTPS.69.160,Tsekov1997,PhysRevE.70.036120}.

\section{Conclusion}\label{conclusion}

Beginning with a $2n$ dimensional stochastic differential equation (SDE) system we derived a generalized Klein-Kramers equation.
It is a Fokker-Planck equation (FPE) that can be obtained from the SDE system regardless the type
of stochastic calculus used (It\^o and Stratonovich integrations of this SDE are described by the same FPE).
This provides a natural starting point to demonstrate the internal consistency of our approach.
After taking the zero mass limit, the $2n$ dimensional SDE reduces to an $n$ dimensional structured form
that is equivalent to the $n$ dimensional Langevin equation with multiplicative noise based on our assumption.
The corresponding generalized Klein-Kramers equation turns to a new FPE.
Such limiting process defines a new type of stochastic integration (A-type) for Langevin equations different from traditional ones.
During the demonstration of consistency, the It\^o \textit{vs.} Stratonovich dilemma is not encountered.
In addition, the new FPE is generally not reachable by the $\alpha$-type integration in higher dimensions.
The framework and the A-type stochastic integration are consequences under a physical view of time evolution dynamics in the phase space.
An attractive advantage of A-type integration is the natural correspondence between stochastic and deterministic dynamics.
For example, fixed points are not changed, which is not possessed by It\^o or Stratonovich's.
As an illustration, experimental data demonstrates that a one-dimensional physical process realizes the zero mass limit and
chooses A-type integration.

\section*{Acknowledgements}
The authors would like to express their sincere gratitude for the inspiring discussions during the weekly group meetings on stochastic processes in biology and related fields;
their appreciation for the valuable comments from an anonymous referee;
and their apology for not being able to include a full list of previous works in the present brief review of a specific approach.
This work was supported in part by the National 973 Project No.~2010CB529200 (P.A.);
and by the Natural Science Foundation of China No.~NFSC91029738 (P.A.) and No.~NFSC61073087 (R.Y.).

\bibliography{Yrs}

\end{document}